# Observation of Momentum-Band Topology in PT-Symmetric Acoustic Floquet Lattices

Shuaishuai Tong,[1] Qicheng Zhang,[1] Gaohan Li,[1] Kun Zhang,[1] Chun Xie,[1] and Chunyin Qiu[1,2*]

[1] *Key Laboratory of Artificial Micro- and Nano-Structures of Ministry of Education and School of Physics and Technology, Wuhan University, Wuhan 430072, China*

[2] *Wuhan Institute of Quantum Technology, Wuhan 430206, China*
[*]*To whom correspondence should be addressed: cyqiu@whu.edu.cn*

*Abstract.* Momentum-band topology—which transcends conventional topological band theory—unlocks new topological phases that host fascinating temporal interface states. However, direct bulk experimental evidence of such emerging band topology is still lacking due to the great challenges in resolving eigenstates and topological invariants of time-varying systems. Here, we present a comprehensive study on the momentum-band topology in a PT-symmetric Floquet lattice, where the drive-induced momentum gap can be characterized by a quantized Berry phase in the energy Brillouin zone. Experimentally, we synthesize the Floquet lattice model using an acoustic cavity-tube structure coupled to custom-designed external circuits. By reconstructing the effective Hamiltonian, we extract the system's eigenstates and provide the first bulk evidence of momentum-band topology from the perspectives of band inversion and topological invariants. This is accompanied by an unambiguous observation of time-localized interface states in real physical time, thereby providing the boundary signature of the bulk topology. Our work paves the way for further experimental studies on the burgeoning momentum-gap physics.

*Introduction*—Over the past decade, topological phases of matter have sparked widespread interest from condensed matter physics [1-4] to diverse classical wave systems [5-10]. In this context, bulk-boundary correspondence plays a central role by revealing the profound connection between the bulk energy-band topology and the existence of boundary states [1,11], as shown in Fig. 1(a). The bulk topology can be characterized by geometric invariants that capture the global property of bulk eigenstates across the momentum Brillouin zone (MBZ), such as the prominent Zak phase [12] and Chern number [13,14] for one- and two-dimensional systems. Furthermore, the introduction of periodic driving has led to the emergence of additional energy Brillouin zone (EBZ) and a series of novel non-equilibrium topological phases [15-24], including anomalous Floquet topological insulators [18,19] and topological space-time crystals with intertwined space-time symmetries [21-24]. Nevertheless, these studies primarily focus on the energy-band topology over the MBZ, with time modulation serving as an extra degree of freedom to control spatial topological interface modes (TIMs).

Momentum-band topology—first proposed in photonic time crystals [25], a type of dielectric media with refractive indices that undergo large, ultrafast periodic variations [25-30]—profoundly advances our understanding of temporal phases of matter. In this context, momentum gaps are formed by the interference of temporal refractions and reflections arising from time-periodic modulations [31-33], which showcase many intriguing properties, such as non-resonant lasers [29,34], subluminal Cherenkov radiation [35], and superluminal momentum gap solitons [36]. In particular, similar to conventional topological phases of matter, the global momentum-band topology in the EBZ results in temporally localized TIMs within momentum gaps [Fig. 1(b)] [25,37]. Although the realization of spatially homogeneous topological photonic time crystals remains a formidable challenge, the concepts of momentum-band topology and resultant time-domain TIMs have been extended to experimentally more accessible spatially discrete systems [38-41]. For example, temporal TIMs have been observed in non-Hermitian optical synthetic lattices [38,39] and static mechanical spatial lattices [41]. However, while the existence of temporal TIMs is linked to the non-trivial momentum-band topology, direct evidence of the bulk topology has not yet been achieved, due to the significant challenges in experimentally characterizing quasi-energy spectra, eigenstates, and topological invariants.

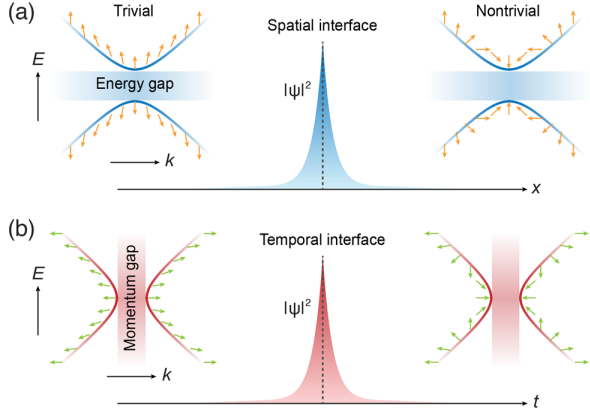

FIG. 1. Schematic of spatial and temporal topological states. (a) Energy-band topology enabling spatial topological interface modes (TIMs). (b) Momentum-band topology enabling temporal TIMs. The evolutions of eigenstates, characterized by arrows, illustrate either trivial or nontrivial band topologies.

In this work, we demonstrate the first bulk evidence of the momentum-band topology in a PT-symmetric Floquet lattice. Theoretically, a nontrivial momentum gap is induced by periodically driven gain and loss, and the momentum-band topology can be characterized by a quantized Berry phase defined in the EBZ. Experimentally, we emulate the Floquet lattice in an acoustic cavity-tube structure incorporated with active circuits, where the dynamic gain/loss configuration is realized by positive/negative feedback circuits while the lattice momentum is synthesized by the phase shift of coupling circuits. By reconstructing the effective Hamiltonian of our acoustic Floquet lattice, we extract both the band structure and eigenstates, and provide direct evidence for the nontrivial momentum-band topology through the examination of band inversion and bulk topological invariants. As complementary evidence from the perspective of temporal bulk-boundary correspondence, we further observe time-domain TIMs within the Floquet momentum gap, which exhibit sound intensity localization at a nontrivial temporal interface. Our findings offer new insights into the novel phenomena associated with momentum gaps in the EBZ and have fundamental implications for the field of nonequilibrium topological physics.

*Theoretical model*—As illustrated in Fig. 2(a), we consider a PT-symmetric Floquet lattice, where the two atoms in each unit cell are coupled via nearest-neighbor hopping $w$ and experience balanced gain and loss $\pm[\gamma_s + \gamma_d(t)]$. Here, $\pm\gamma_s$ denote static gain/loss, while $\pm\gamma_d(t) = \pm\gamma S(t)$ represent dynamic gain/loss, where $S(t) = \text{sgn}[\cos(2\pi\Omega t)]$ is a square-wave function with time period $T = 1/\Omega$. The momentum-space Hamiltonian of this periodically driven lattice reads

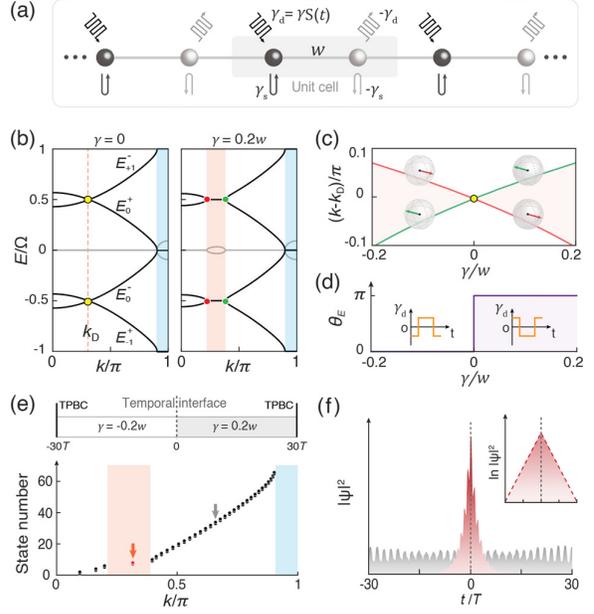

FIG. 2. Topological momentum gap in a PT-symmetric Floquet lattice. (a) Tight-binding model. It features nearest neighboring hopping $w$, static gain/loss $\pm\gamma_s$, and dynamic gain/loss $\pm\gamma_d(t) = \pm\gamma S(t)$, with $S(t)$ being a $T$-period square wave function. (b) Band structures for the lattices without (left) and with (right) dynamic gain/loss applied, where the dark and grey lines represent the real and imaginary parts of the energy, respectively. The nonzero dynamic gain/loss splits the Floquet Dirac points (yellow dots) into a pair of Floquet EPs (red and green dots). (c) $\gamma$-dependence of the Floquet-EP momenta, indicating a band inversion in momentum. Insets: typical Floquet-EP states sketched on the Bloch sphere. (d) Berry phase $\theta_E$ plotted as a function of $\gamma$, where the insets illustrate the profiles of dynamic gain/loss for $\gamma < 0$ and $\gamma > 0$. (e) Momentum spectrum (lower panel) of a temporal domain-wall structure (upper panel) with temporal periodic boundary condition (TPBC). (f) Probability density distributions of the bulk mode (gray) and TIM (red) across the domain-wall structure. The latter is well captured by the simplified low-energy model around Floquet Dirac point (red dashed line, inset).

$$\mathbf{H}(t) = \begin{bmatrix} i\gamma_s + i\gamma_d(t) & w + we^{ik} \\ w + we^{-ik} & -i\gamma_s - i\gamma_d(t) \end{bmatrix}. \quad (1)$$

The quasi-energy spectrum can be solved from the effective Hamiltonian $\mathbf{H}_{\text{eff}} = (i2\pi T)^{-1}\ln\mathbf{U}$, where the Floquet operator $\mathbf{U} = e^{-i2\pi\mathcal{T}\int_0^T \mathbf{H}(t)dt}$ with $\mathcal{T}$ being time-ordering operator. Without loss of generality, we consider $\Omega = 3.5w$, $\gamma_s = 0.3w$, and $\gamma = 0.2w$. Figure 2(b) presents the band structure for the system (right panel), compared to the case without dynamic modulation (left panel). For clarity, we provide the quasi-energy bands within two EBZs and half of the MBZ [notice that $E(-k) = E(k)$]. Clearly, it shows that the virtual Dirac points at the EBZ boundaries, which are crossed by neighboring quasi-energy bands at $\gamma = 0$, split into pairwise Floquet EPs and form Floquet momentum gaps due to the non-zero dynamic modulation. Note that the momentum gap around $k = \pi$, referred to as the static momentum gap, is induced by a non-zero $\gamma_s$.

Now, we investigate the topology of the momentum gap induced by the dynamic modulation of gain/loss. To do this, we first examine the evolution of the Floquet-EP states as a function of the parameter $\gamma$, which characterizes the dynamic gain/loss in our PT-symmetric Floquet lattices. As shown in Fig. 2(c), the $\gamma$-dependence of the Floquet-EP momenta reveals that the two EP states swap their momentum order when crossing the virtual Dirac point at $\gamma = 0$. Similar to the energy-band inversion in conventional topological band theory [1,42], the momentum-band inversion marks a signature of topological transition. In our PT-symmetric Floquet system, the momentum-band topology can be characterized by a quantized Berry phase defined in the EBZ

$$\theta_E = i \oint_{EBZ} \psi^\dagger(E) \partial_E \psi(E) dE, \quad (2)$$

with $\psi(E)$ being the eigenstate of the effective Hamiltonian $\mathbf{H}_{\text{eff}}$. Figure 2(d) presents the $\gamma$-dependent Berry phase calculated for the momentum band between the static and Floquet momentum gaps. It shows two topologically distinct phases: $\theta_E = 0$ for $\gamma < 0$ and $\theta_E = \pi$ for $\gamma > 0$. According to the bulk-boundary correspondence, one can expect temporally localized states at a time interface formed by the two phases. To demonstrate this, as shown in Fig. 2(e), we construct a temporal domain-wall structure composed of the Floquet lattices with $\gamma = \pm 0.2w$, and calculate the momentum spectrum under a time-periodic boundary condition (see Supplemental Material [43] for details). The spectrum reveals isolated topological states (red dot) within the Floquet momentum gap. Accordingly, the time-domain probability density distributions in Fig. 2(f) demonstrate that the in-gap TIM (red area) is strongly localized around the time interface, in sharp contrast to the temporally extended bulk state (grey area). Note that the exponential decay of the TIM can be precisely captured by a low-energy model with a momentum operator $M_D(\delta E) = v_D^{-1}(m\sigma_y - \delta E \sigma_z)$— a momentum version of the effective Dirac Hamiltonian. Here $\delta E$ represents the energy deviation from the Floquet Dirac point, $v_D$ is Dirac velocity, and $\sigma_{y,z}$ are Pauli matrices, respectively. Resembling the conventional Dirac model that characterizes the energy-gap topology, the momentum-gap topology depends on the sign of Dirac mass $m$, which is proportional to $\gamma$. As the consequence, an interface separating the time domains with Dirac masses of opposite signs supports nontrivial temporal TIMs. (see Supplemental Material [43]).

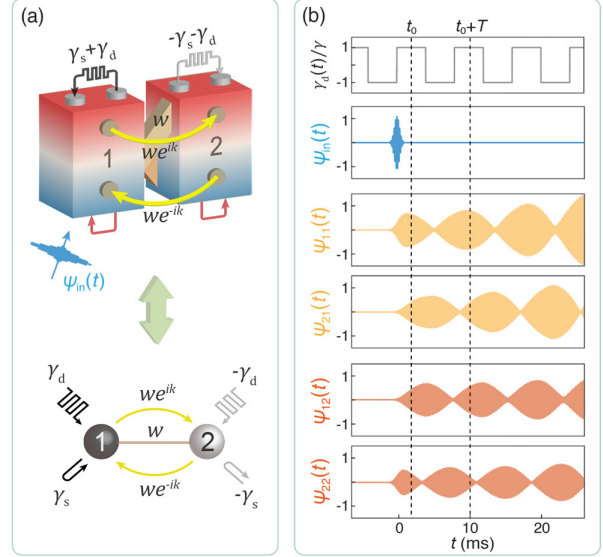

FIG. 3. Acoustic emulation of the PT-symmetric Floquet lattice in synthetic space. (a) Schematic diagram of our experimental setup. The acoustic cavities mimic the atoms and the narrow tubes between them emulate the reciprocal intracell hopping, respectively. Positive and negative feedback circuits, illustrated by black and grey arrows, provide desired gain and loss configuration, while the phase-controllable coupling circuits (yellow arrows) enable the intercell hopping $we^{ik}$. (b) Measured time-domain wavefunctions $\boldsymbol{\psi}_1(t) = (\psi_{11}, \psi_{21})^T$ and $\boldsymbol{\psi}_2(t) = (\psi_{12}, \psi_{22})^T$, which are generated by injecting a Gaussian sound signal $\psi_{\text{in}}(t)$ into cavities 1 and 2, respectively. The sound signals within a $T$-period are extracted for reconstructing the effective Hamiltonian. The experimental data, exemplified by $k = 0.3\pi$, correspond to the system with $w = 36$ Hz, $\gamma_s = 11$ Hz, $\gamma = 7$ Hz, and $\Omega = 125$ Hz.

*Acoustic implementation of PT-symmetric Floquet lattices*—Experimentally, we simulate the PT-symmetric Floquet lattice utilizing a passive acoustic cavity-tube structure supplied with active circuits. As sketched in Fig. 3(a), the air-filled cavities emulate the lattice sites with onsite energy $f_0 = 5100$ Hz, while the narrow tubes between them provide static and reciprocal intracell coupling $w = 36$ Hz [44-46]. To implement the desired gain/loss configuration, we utilize a set of time-dependent positive feedback circuits to provide the gain $\gamma_s + \gamma_d(t)$ in cavity 1, while using a set of negative feedback circuits to introduce the balanced loss $-\gamma_s - \gamma_d(t)$ in cavity 2. The temporal modulation is synchronized by a square-wave voltage generated by a waveform generator [47-50]. In addition, two sets of unidirectional coupling circuits are introduced to mimic the intercell hoppings $we^{\pm ik}$. More details about the experimental setup are provided in the Supplemental Material [43]. Ultimately, we synthesize the PT-symmetric Floquet lattice governed by the $k$-dependent

Hamiltonian in Eq. (1). Note that our experimental setup allows flexible momentum selection by simply regulating the phases of coupling circuits. This facilitates a precise measurement of the energy-momentum spectrum and eigenstates, thereby enabling a direct characterization of the momentum-band topology in the EBZ.

To experimentally extract the energy-momentum spectra as well as eigenstates of our acoustic systems, we propose a simple approach to reconstruct the effective Hamiltonian $\mathbf{H}_{\text{eff}}$. As shown in Fig. 3(b), we inject a pulsed acoustic signal $\psi_{\text{in}}(t)$ into cavity $i$, and measure the resultant sound response in cavity $j$, denoted by $\psi_{ji}(t)$ with $i,j = 1,2$. This gives two linearly independent wavefunctions $\boldsymbol{\psi}_1(t) = (\psi_{11}, \psi_{21})^{\text{T}}$ and $\boldsymbol{\psi}_2(t) = (\psi_{12}, \psi_{22})^{\text{T}}$. Theoretically, for the wavefunction $\boldsymbol{\psi}_{1,2}(t)$ that evolves freely over time, $\boldsymbol{\psi}_{1,2}(t_0)$ and $\boldsymbol{\psi}_{1,2}(t_0 + T)$ are related by the Floquet operator, $\boldsymbol{\psi}_{1,2}(t_0 + T) = \mathbf{U}\boldsymbol{\psi}_{1,2}(t_0)$. Consequently, using the two experimentally measured states, we reconstruct the $2 \times 2$ Floquet operator via
$$\mathbf{U} = [\boldsymbol{\psi}_1(t_0 + T), \boldsymbol{\psi}_2(t_0 + T)][\boldsymbol{\psi}_1(t_0), \boldsymbol{\psi}_2(t_0)]^{-1}, \quad (3)$$
which further gives the effective Hamiltonian of the Floquet lattice: $\mathbf{H}_{\text{eff}} = (i2\pi T)^{-1}\ln \mathbf{U}$. More specifically, as shown in Fig. 3(b), we consider a short-duration Gaussian wavepacket $\psi_{\text{in}}(t) = \exp(-t^2/\Delta t^2)\exp(i2\pi f_0 t)$ with $\Delta t = 0.5 \text{ ms}$, and set $t_0 = 2 \text{ ms}$ to ensure sufficient temporal separation between the resultant sound responses over a single $T$-period and the incident pulse signal, since $\psi_{\text{in}}(t) \to 0$ when $t > t_0$. Based on the reconstructed effective Hamiltonian $\mathbf{H}_{\text{eff}}(k)$, we obtain the quasi-energies and eigenstates of the PT-symmetric Floquet lattice for each momentum, as well as the energy-momentum spectrum by sweeping the phase of coupling circuits over $2\pi$.

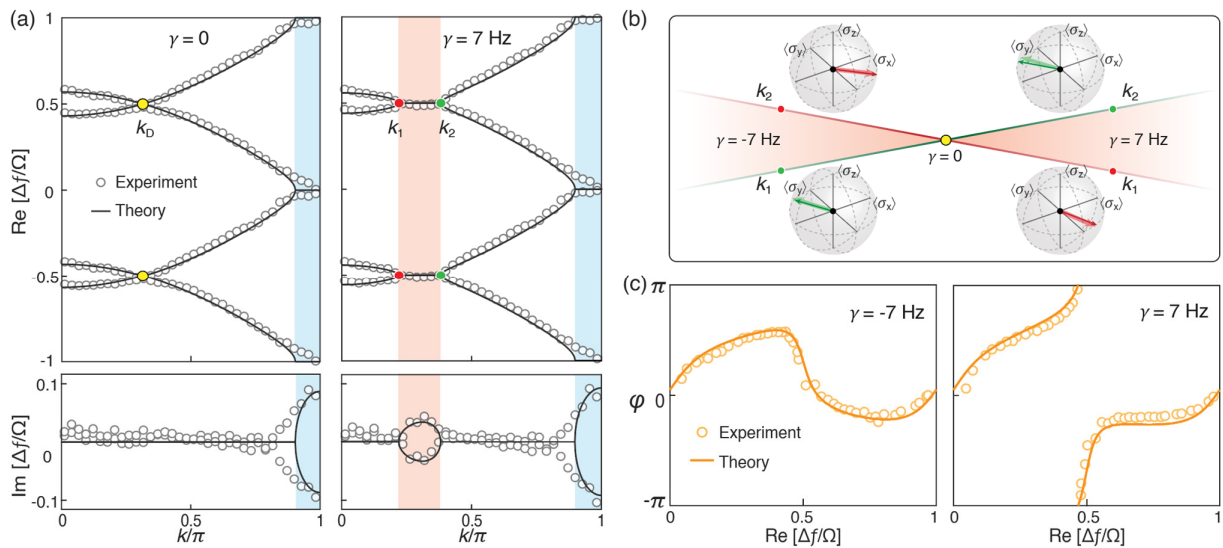

FIG. 4. Experimental evidence for momentum-band topology in the EBZ. (a) Comparative energy-momentum spectra measured for the systems with $\gamma = 0$ (left) and $\gamma = 7 \text{ Hz}$ (right), respectively. For clarity, here the real and imaginary spectra are plotted separately, and the frequency $\Delta f$ is measured from the resonant frequency of single cavity. The blue shaded area labels the normal momentum gap induced by the static gain/loss, while the orange shaded area highlights the Floquet momentum gap driven by the dynamic modulation. (b) Floquet-EP states (thick arrows) measured for the systems with $\gamma = -7 \text{ Hz}$ (left) and $\gamma = 7 \text{ Hz}$ (right), compared with the simulation results (thin arrows). (c) Phase evolutions of PT-eigenvalues within the EBZ for the systems of $\gamma = -7 \text{ Hz}$ and $\gamma = 7 \text{ Hz}$, demonstrating the trivial and nontrivial momentum-band topologies, respectively. The experimental data (circles) agree well with theoretical results (solid lines).

*Experimental evidence for momentum-band topology—* Figure 4(a) presents the complex energy-momentum spectrum measured for the acoustic system with $\gamma = 7 \text{ Hz}$, in comparison with that of the static case ($\gamma = 0$). It shows that while both systems exhibit a normal momentum gap around $k = \pi$, the presence of dynamic gain/loss opens a momentum gap at the virtual Floquet Dirac point near $k = 0.3\pi$, accompanied by complex-valued energies between the Floquet-EP momenta $k_1 \simeq 0.21\pi$ and $k_2 \simeq 0.39\pi$. The presence of the Floquet momentum gap is further corroborated by the observation of exponentially growing sound intensity over time, as detailed in the Appendix.

Below, we directly demonstrate the momentum-band topology in the EBZ by analyzing the eigenstates of our experimentally reconstructed effective Hamiltonian. Figure 4(b) shows the measured Floquet-EP states for the systems with $\gamma = \pm 7 \text{ Hz}$, indicated by thick arrows on the Bloch spheres, which match well the theoretical predictions (thin arrows). [Note that the experimental energy-momentum spectrum of $\gamma = -7$ Hz, not presented in Fig. 4(a), is nearly indistinguishable from that of $\gamma = 7$ Hz.] As predicted, the two Floquet-EP states exchange their momenta ($k_1$ and $k_2$) when the sign of $\gamma$ is reversed—an essential signature of the momentum-band inversion in our acoustic Floquet lattice. Furthermore, we characterize the Berry phase based on the experimentally measured eigenstates. Notably, due to the PT-symmetric nature of the real-energy eigenstates in between the static and Floquet momentum gaps, the Berry phase can be equivalently expressed through the winding angle of the PT eigenvalue $e^{i\varphi(E)} = \psi^\dagger(E)\mathbf{PT}\psi(E)$ around the origin of the complex plane (see Supplemental Material [43] for details). This results in an alternative formulation for the Berry phase

$$\theta_E = \frac{1}{2}\oint_{EBZ} \partial_E \varphi(E) dE. \quad (4)$$

This expression greatly simplifies our experimental characterization. Figure 4(c) shows the evolution of measured phase angles $\varphi(E)$ for the systems with $\gamma = \pm 7$ Hz. For the system with $\gamma = -7$ Hz, the phase exhibits no winding across the EBZ, indicating a trivial topology with $\theta_E = 0$. In contrast, for $\gamma = 7$ Hz, the phase winds by $2\pi$, identifying the nontrivial topology with $\theta_E = \pi$. All experimental results (circles) agree excellently with the theoretical predictions (solid lines), providing clear evidence distinguishing the momentum-band topologies of the two systems in the EBZ.

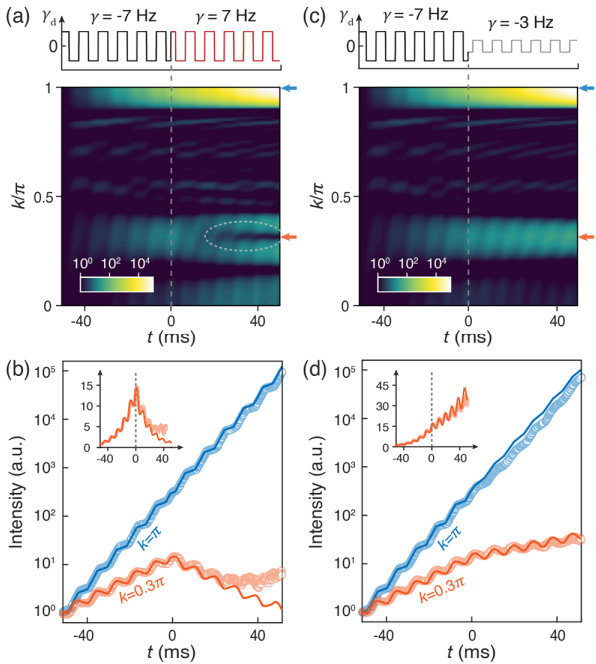

FIG. 5. Experimental observation of acoustic temporal TIMs. (a) Momentum-resolved time evolution of sound intensity simulated for a nontrivial temporal interface system. The red and blue arrows mark the momenta $k = 0.3\pi$ and $k = \pi$, respectively. (b) Sound intensities at these two momenta, with experimental data (circles) closely matching the simulations (lines). A base-10 logarithmic scale is used to better visualize the exponential behavior. The data for $k = 0.3\pi$, along with their linear-scale plot (inset), clearly show the emergence of a TIM at the temporal interface ($t = 0$ ms). (c), (d) Similar to (a) and (b), but for a trivial temporal interface system, showing no TIM at the interface.

*Observation of acoustic temporal TIM*—To demonstrate the intriguing temporal TIM in our acoustic system, we investigate the wave dynamics around the temporal interface at $t = 0$, realized by setting $\gamma = -7$ Hz for $t < 0$ and $\gamma = 7$ Hz for $t > 0$. Figure 5(a) presents the momentum-resolved sound intensity simulated as a function of time, with an initial pulse excitation applied to cavity 1. In contrast to the monotonic signal growth within the static momentum-gap around $k = \pi$, the sound intensity within the Floquet momentum-gap (near $k = 0.3\pi$) manifests a fine structure. Specifically, within an extremely narrow momentum range, the sound signal initially increases but then decreases after the temporal interface—a hallmark characteristic of temporal TIM. More details are provided in Supplemental Material [43]. To experimentally identify these fundamental signatures, we focus on the momenta $k = 0.3\pi$ and $k = \pi$, and switch the gain/loss configuration to form a nontrivial temporal interface by simply controlling the waveform generator. The measured sound intensity for $k = 0.3\pi$ clearly confirms the emergence of the temporally localized TIM [Fig. 5(b)]. The deviation between the experimental and simulated results for $t > 25$ ms arises from the high sensitivity to the synthesized momentum, which ultimately relies on the accuracy of the unidirectional couplings $we^{\pm ik}$. (This is in sharp contrast to the excellent agreement for the case of $k = \pi$, which manifests a stable exponential growth of sound signal.) Furthermore, we present the comparative intensity spectra for a temporal interface composed of two topologically trivial systems with $\gamma = -7$ Hz and $\gamma = -3$ Hz [Figs. 5(c)-5(d)]. As expected, the intensities in both momentum gaps increase monotonically with time and show no temporal TIMs in this case.

*Conclusion and discussion*—We present a comprehensive and in-depth experimental study of the momentum-band topology in a PT-symmetric Floquet lattice. To achieve this, we synthesize the Floquet lattice using an acoustic cavity-tube system equipped with precisely controlled external circuits. We not only provide the first direct bulk evidence of momentum-band topology from the perspective of band-inversion and topological invariants, but also demonstrate the intriguing TIM in real physical time, highlighting the role of temporal bulk-boundary correspondence. The conclusive verification of momentum-band topology could inspire further experimental studies on temporal Thouless pumping in lattice systems, which is a fascinating direction as the causality constraint may lead to conclusions fundamentally distinct from their spatial counterparts [51]. Moreover, the key technique in our experiments—reconstructing the Floquet effective Hamiltonian by measuring the temporal evolution of wavefunctions—can be widely applied to explore highly elusive Floquet non-equilibrium physics. Our acoustic platform, with its excellent tunability and reconfigurability, holds significant promise for investigating a broad range of temporal matter physics. This includes, for example, temporal quasicrystals [51], time-domain Anderson effects [52,53], and temporal Moiré superlattices [54,55].

*Acknowledgements*—This project is supported by the National Key R&D Program of China (Grant No. 2023YFA1406900), the National Natural Science Foundation of China (Grants No. 12374418, No. 12304495, and No. 12104346), the China Postdoctoral Science Foundation (Grant No. 2024M762462), and the Fundamental Research Funds for the Central Universities.

*References*
1. M. Z. Hasan, and C. L. Kane, Colloquium: Topological insulators, Rev. Mod. Phys. **82**, 3045 (2010).
2. X.-L. Qi and S.-C. Zhang, Topological insulators and superconductors, Rev. Mod. Phys. **83**, 1057 (2011).
3. N. P. Armitage, E. J. Mele, and Ashvin Vishwanath, Weyl and Dirac semimetals in three-dimensional solids, Rev. Mod. Phys. **90**, 015001 (2018).
4. M. Zahid Hasan, G. Chang, I. Belopolski, G. Bian, S.-Y. Xu, and J.-X. Yin, Weyl, Dirac and high-fold chiral fermions in topological quantum matter, Nat. Rev. Mater. **6**, 784–803 (2021).
5. L. Lu, J. D. Joannopoulos, and M. Soljačić, Topological photonics. Nat. Photon. **8**, 821–829 (2014).
6. T. Ozawa, H. M. Price, A. Amo, N. Goldman, M. Hafezi, L. Lu, M. C. Rechtsman, D. Schuster, J. Simon, O. Zilberberg, and I. Carusotto, Topological photonics, Rev. Mod. Phys. **91**, 015006 (2019).
7. H. He, C. Qiu, L. Ye, X. Cai, X. Fan, M. Ke, F. Zhang, and Z. Liu, Topological negative refraction of surface acoustic waves in a Weyl phononic crystal, Nature **560**, 61–64 (2018).
8. X. Zhang, F. Zangeneh-Nejad, Z.-G. Chen, M.-H. Lu, and J. Christensen, A second wave of topological phenomena in photonics and acoustics, Nature **618**, 687–697 (2023).
9. V. V. Albert, L. I. Glazman, and L. Jiang, Topological Properties of Linear Circuit Lattices, Phys. Rev. Lett. **114**, 173902 (2015).
10. S. Imhof, C. Berger, F. Bayer, J. Brehm, L. W. Molenkamp, T. Kiessling, F. Schindler, C. H. Lee, M. Greiter, T. Neupert, and Ronny Thomale, Topolectrical-circuit realization of topological corner modes, Nat. Phys. **14**, 925–929 (2018).
11. A. M. Essin, and V. Gurarie, Bulk-boundary correspondence of topological insulators from their respective Green's functions, Phys. Rev. B **84**, 125132 (2011).
12. J. Zak, Berry's phase for energy bands in solids, Phys. Rev. Lett. **62**, 2747 (1989).
13. D. J. Thouless, M. Kohmoto, M. P. Nightingale, and M. den Nijs, Quantized Hall Conductance in a Two-Dimensional Periodic Potential, Phys. Rev. Lett. **49**, 405 (1982).
14. B. Simon, Holonomy, the Quantum Adiabatic Theorem, and Berry's Phase, Phys. Rev. Lett. **51**, 2167 (1983).
15. N. Lindner, G. Refael, and V. Galitski, Floquet topological insulator in semiconductor quantum wells. Nat. Phys. **7**, 490–495 (2011).
16. M. C. Rechtsman, J. M. Zeuner, Y. Plotnik, Y. Lumer, D. Podolsky, F. Dreisow, S. Nolte, M. Segev and A. Szameit, Photonic Floquet topological insulators, Nature **496**, 196–200 (2013).
17. Á. Gómez-León, and G. Platero, Floquet-Bloch Theory and Topology in Periodically Driven Lattices, Phys. Rev. Lett. **110**, 200403 (2013).
18. M. S. Rudner, N. H. Lindner, E. Berg, and M. Levin, Anomalous Edge States and the Bulk-Edge Correspondence for Periodically Driven Two-Dimensional Systems, Phys. Rev. X **3**, 031005 (2013).
19. L. J. Maczewsky, J. M. Zeuner, S. Nolte, and A. Szameit, Observation of photonic anomalous Floquet topological insulators. Nat. Commun. **8**, 13756 (2017).
20. M. S. Rudner, and N. H. Lindner, Band structure engineering and non-equilibrium dynamics in Floquet topological insulators. Nat. Rev. Phys. **2**, 229–244 (2020).
21. T. Morimoto, H. Chun Po, and A. Vishwanath, Floquet topological phases protected by time glide symmetry, Phys. Rev. B **95**, 195155 (2017).
22. Y. Peng and G. Refael, Floquet second-order topological insulators from nonsymmorphic space-time symmetries, Phys. Rev. Lett. **123**, 016806 (2019).
23. J. Jin, L. He, J. Lu, E. J. Mele, and B. Zhen Floquet Quadrupole Photonic Crystals Protected by Space-Time Symmetry. Phys. Rev. Lett. **129**, 063902 (2022).
24. Y. Peng, Topological Space-Time Crystal, Phys. Rev. Lett. **128**. 186802 (2022).
25. E. Lustig, Y. Sharabi, and M. Segev, Topological aspects of photonic time crystals, Optica **5**, 1390-1395 (2018).
26. F. Biancalana, A. Amann, A. V. Uskov, and E. P. O'Reilly, Dynamics of light propagation in spatiotemporal dielectric structures, Phys. Rev. E **75**, 046607 (2007).
27. J. R. Zurita-Sánchez, P. Halevi, and J. C. Cervantes-González, Reflection and transmission of a wave incident on a slab with a time-periodic dielectric function $\epsilon(t)$, Phys. Rev. A **79**, 053821 (2009).
28. N. Wang, Z.-Q. Zhang, and C. T. Chan, Photonic Floquet media with a complex time-periodic permittivity, Phys. Rev. B **98**, 085142 (2018).
29. M. Lyubarov, Y. Lumer, A. Dikopoltsev, E. Lustig, Y. Sharabi, and M. Segev, Amplified emission and lasing in photonic time crystals, Science **377**, 425–428 (2022).
30. X. Wang, P. Garg, M. S. Mirmoosa, A. G. Lamprianidis, C. Rockstuhl, and V. S. Asadchy, Expanding momentum bandgaps in photonic time crystals through resonances. Nat. Photon. **19**, 149–155 (2025).
31. B. Apffel, and E. Fort, Frequency Conversion Cascade by Crossing Multiple Space and Time Interfaces, Phys. Rev. Lett. **128**, 064501 (2022).
32. J. R. Reyes-Ayona, P. Halevi, Observation of genuine wave vector ($k$ or $\beta$) gap in a dynamic transmission line and temporal photonic crystals, Appl. Phys. Lett. **107**, 074101 (2015).
33. H. Moussa, G. Xu, S. Yin, E. Galiffi, Y.s Ra'di, and Andrea Alù, Observation of temporal reflection and broadband frequency translation at photonic time interfaces. Nat. Phys. **19**, 863–868 (2023).
34. H. Li, S. Yin, H. He, J. Xu, A. Alù, and B. Shapiro, Stationary Charge Radiation in Anisotropic Photonic Time Crystals, Phys. Rev. Lett. **130**, 093803 (2023).

35. A. Dikopoltsev, Y. Sharabi, M. Lyubarov, Y. Lumer, S. Tsesses, E. Lustig, I. Kaminer, and M. Segev, Light emission by free electrons in photonic time-crystals, Proc. Natl. Acad. Sci. U.S.A. **119**, e2119705119 (2022).
36. Y. Pan, M.-I. Cohen, and M. Segev, Superluminal *k*-Gap Solitons in Nonlinear Photonic Time Crystals, Phys. Rev. Lett. **130**, 233801 (2023).
37. Y. Yang, H. Hu, L. Liu, Y. Yang, Y. Yu, Y. Long, X. Zheng, Y. Luo, Z. Li, and F. J. Garcia-Vidal, Topologically Protected Edge States in Time Photonic Crystals with Chiral Symmetry, ACS Photonics (2025).
38. Y. Ren, K. Ye, Q. Chen, F. Chen, L. Zhang, Y. Pan, W. Li, X. Li, L. Zhang, H. Chen, and Y. Yang, Observation of momentum-gap topology of light at temporal interfaces in a time-synthetic lattice. Nat. Commun. **16**, 707 (2025).
39. J. Feis, S. Weidemann, T. Sheppard, H. M. Price, and A. Szameit, Space-time-topological events in photonic quantum walks. Nat. Photon. **19**, 518–525 (2025).
40. W. Zhu, and J.-H. Jiang, Characterizing generalized Floquet topological states in hybrid space-time dimensions, arXiv:2409.09937 (2024).
41. M. Li, J. Liu, X. Wang, W. Chen, G. Ma, and J. Dong, Topological temporal boundary states in a non-Hermitian spatial crystal, arXiv:2306.09627v2 (2024).
42. B. A. Bernevig, T.L. Hughes, and S.-C. Zhang, Quantum Spin Hall Effect and Topological Phase Transition in HgTe Quantum Wells, Science **314**, 1757 (2006).
43. See Supplemental Material [url] for theoretical details and experimental setup, which includes Refs. [38,48,49,56,57].
44. H. Xue, Y. Yang, F. Gao, Y. Chong, B. Zhang, Acoustic higher-order topological insulator on a kagome lattice, Nat. Mater. **18**, 108–112 (2019).
45. X. Ni, M. Weiner, A. Alù, and A. B. Khanikaev, Observation of higher-order topological acoustic states protected by generalized chiral symmetry, Nat. Mater. **18**, 113–120 (2019).
46. Y. Qi, C. Qiu, M. Xiao, H. He, M. Ke, and Z. Liu, Acoustic Realization of Quadrupole Topological Insulators, Phys. Rev. Lett. **124**, 206601 (2020).
47. L. Zhang, Y. Yang, Y. Ge, Y. Guan, Q. Chen, Q. Yan, F. Chen, R. Xi, Y. Li, D. Jia, S. Yuan, H. Sun, H. Chen, and B. Zhang, Acoustic non-Hermitian skin effect from twisted winding topology. Nat. Commun. **12**, 6297 (2021).
48. J. Liu, Z. Li, Z.-G. Chen, W. Tang, A. Chen, B. Liang, G. Ma, and J.-C. Cheng, Experimental Realization of Weyl Exceptional Rings in a Synthetic Three-Dimensional Non-Hermitian Phononic Crystal, Phys. Rev. Lett. **129**, 084301 (2022).
49. Q. Zhang, Y. Li, H. Sun, X. Liu, L. Zhao, X. Feng, X. Fan, and C. Qiu, Observation of Acoustic Non-Hermitian Bloch Braids and Associated Topological Phase Transitions, Phys. Rev. Lett. **130**, 017201 (2023).
50. Z. Chen, A. Chen, Y.-G. Peng, Z. Li, B. Liang, J. Yang, X.-F. Zhu, Y. Lu, and J. Cheng, Observation of acoustic Floquet π modes in a time-varying lattice, Phys. Rev. B **109**, L020302 (2024).
51. X. Ni, S. Yin, H. Li, and A. Alù, Topological wave phenomena in photonic time quasicrystals, Phys. Rev. B **111**, 125421 (2025).
52. R. Carminati, H. Chen, R. Pierrat, and B. Shapiro, Universal Statistics of Waves in a Random Time-Varying Medium, Phys. Rev. Lett. **127**, 094101 (2021).
53. Y. Sharabi, E. Lustig, and M. Segev, Disordered Photonic Time Crystals, Phys. Rev. Lett. **126**, 163902 (2021).
54. L. Zou, H. Hu, H. Wu, Y. Long, Y. Chong, B. Zhang, Y. Luo, Momentum flatband and superluminal propagation in a photonic time Moiré superlattice, arXiv:2411.00215 (2024).
55. Z. Dong, X. Chen, L. Yuan, Extreme narrow band in Moiré Photonic time crystal, arXiv:2411.05351 (2024).
56. R. de Gail1, M. O. Goerbig, F. Guinea, G. Montambaux, and A. H. Castro Neto, Topologically protected zero modes in twisted bilayer graphene, Phys. Rev. B 84, 045436 (2011).
57. A. Stegmaier, S. Imhof, T. Helbig, T. Hofmann, C. H. Lee, M. Kremer, A. Fritzsche, T. Feichtner, S. Klembt, et al, Topological Defect Engineering and Symmetry in Non-Hermitian Electrical Circuits, Phys. Rev. Lett. **126**, 215302 (2021).